\documentclass[prl, twocolumn, amsmath, amssymb,superscriptaddress]{revtex4}

\usepackage{graphicx,color,amsmath,amssymb,bm,hyperref} 
          
\newcommand{\tr}{\mathop{\mathrm{tr}}\nolimits}     
    \newcommand{\SA}{\mathop{\mathcal{S}}\nolimits}
     
\newcommand{\I}{\mathop{\mathbb{I}}\nolimits}   

\newcommand{\ketbra}[2]{| #1 \rangle \langle #2 |}

\newcommand{\ket}[1]{| #1 \rangle}

\newcommand{\EA}{\mathop{\mathcal{E}}\nolimits}

\begin{document}
\date{\today}
\title{Lazy states: sufficient and necessary condition for zero quantum entropy rates under any coupling to the environment}
  
\author{C\'{e}sar A. Rodr\'{i}guez-Rosario} \email[email:
]{rodriguez[at]chemistry.harvard.edu}  \affiliation{  Department of Chemistry and Chemical Biology,\\ Harvard
  University, Cambridge, Massachusetts 02138, USA }

\author{Gen Kimura}
\email[email: ]{gen-kimura[at]aist.go.jp} \affiliation{Research Center for Information Security (RCIS),
National Institute of Advanced Industrial
Science and Technology (AIST).
Daibiru building 1003,
1-18-13 Sotokanda, Chiyoda-ku, Tokyo, 101-0021, Japan}

\author{Hideki Imai}
\affiliation{Research Center for Information Security (RCIS),
National Institute of Advanced Industrial
Science and Technology (AIST).
Daibiru building 1003,
1-18-13 Sotokanda, Chiyoda-ku, Tokyo, 101-0021, Japan}
\affiliation{Graduate School of Science and Engineering,
Chuo University.
1-13-27 Kasuga, Bunkyo-ku, Tokyo 112-8551, Japan}

\author{Al\'{a}n
  Aspuru-Guzik} \email[email: ]{aspuru[at]chemistry.harvard.edu}
\affiliation{  Department of Chemistry and Chemical Biology,\\ Harvard
  University, Cambridge, Massachusetts 02138, USA }

\begin{abstract}
We find the necessary and sufficient conditions for the entropy rate of the system to be zero under any system-environment Hamiltonian interaction. We call the class of system-environment states that satisfy this condition lazy states. They are a generalization of classically correlated states defined by quantum discord, but based on projective measurements of any rank. The concept of lazy states permits the construction of a protocol for detecting global quantum correlations using only  local dynamical information. We show how quantum correlations to the environment provide bounds to the entropy rate, and how to estimate dissipation rates for general non-Markovian open quantum systems.
\end{abstract}

\pacs{03.65.Xp,03.65.Yz,03.67.-a,05.70.-a} 
\keywords{open systems, positive maps}

\maketitle 

Protecting a quantum system from decoherence is one of the main challenges in quantum computation \cite{RevModPhys.75.715}. It is common to only consider a system state $(\mathcal{S})$ initially uncorrelated from its environment $(\mathcal{E})$, but that is weakly coupled to it. A measure that quantifies the degree of decoherence of the system is the von Neumann entropy $\mathbb{S}:= -\tr\left\{ \rho^\mathcal{S} \ln \rho^\mathcal{S} \right\}$. In such a case, decoherence arises from correlations developed as a consequence of the $\mathcal{SE}$ coupling which change the system entropy. Starting from this point of view that assumes that the system undergoes open quantum system dynamics \cite{PhysRev.121.920,Gorini76a,Rodriguez08}, the entropy of quantum systems can be preserved from decoherence. Symmetries of the dynamics can be exploited \cite{Zanardi97,Zanardi98,PhysRevLett.81.2594,Zanardi04,Roa07} or dynamical operations can be performed \cite{PhysRevLett.89.080401,Zanardi99, PhysRevLett.82.2417} to stabilize the entropy of a system.

We depart from that approach that focuses on the structure of the open system dynamics with an unknown environment and the standard approximations that go with it. Instead, we take the point of view of studying the structure of the total system-environment states $\rho^\mathcal{SE}$ and their relationship to the system decoherence. This allows us to make theoretical progress, without resorting to assuming explicit knowledge of the $\mathcal{SE}$ Hamiltonian. In this paper, we find universal properties of the system entropy rates that are independent of the details of the $\mathcal{SE}$ coupling. By taking this different approach, we find the necessary and sufficient conditions for the class of system-environment states for which the rate of change of the system entropy at a time $\tau$ (entropy rate) is zero, $\frac{d}{dt}\mathbb{S}(\rho^\mathcal{S}) \vert_{t=\tau}$, for \emph{any} type of interaction with the environment $H_{int}$. 

We also demonstrate how correlations to the environment not only lead to decoherence, but also provide a bound on the decoherence rate valid beyond the weak-coupling limit and without any assumptions of Markovianity. Our proof goes beyond previous work by Kimura~\emph{et al. }\cite{ref:GOH,ref:GOM} which showed that $\mathcal{SE}$ correlations were a necessary condition for the change of purity (linear entropy) under any interaction. 

To prove these results, we consider general evolution of the density operator in finite-dimensional Hilbert space $\mathcal{SE}$ given by the von Neumann equation, $\frac{d}{dt} \rho^\mathcal{SE}_t\,\vert_{t=\tau} = -i \left[H_{tot},\,\rho^\mathcal{SE}_\tau\right]$. The total Hamiltonian is $H_{tot} = H^\mathcal{S}\otimes \I^\mathcal{E} + \I^\mathcal{S} \otimes H^\mathcal{E} + H_{int}$ \footnote{$H_{int}$ is such that $\tr_\mathcal{S}\left\{ H_{int} \right\}=\tr_\mathcal{E}\left\{ H_{int} \right\}=0$.}, which consists of the system, the environment and the interaction  Hamiltonians \footnote{We assume that $H_{tot}$ is a bounded operator. Similar results can be generalized to an unbounded Hamiltonian by considering an operator domain and imposing the assumption of the finite variance of energy, as shown in \cite{ref:GOM}.}. The time dependence of $H_{tot}$ is implicit, but without loss of generality we write $H_{tot}$ in the picture where $H_{int}$ is time-independent. Using this, we want to consider properties of the dynamics of the system density matrix at time $\tau$, $\rho^\mathcal{S}_\tau=\sum_j p_j \Pi_j^\mathcal{S}$, where $\left\{ \Pi_j \right\}$ are orthonormal projectors of any rank chosen such that $\{p_j\}$ is non-degenerate. 

We start by considering functions of the form of $f_N( \rho^\mathcal{S} )=\tr_\mathcal{S}\left\{ (\rho^\mathcal{S})^N\right\}=\sum_j p_j^N$. By taking the trace of the environment in the von Neumann equation, and taking the time derivative of $f_N$, we obtain, $\left[ \frac{d}{dt}f_N( \rho^\mathcal{S}_t )\right]_{t=\tau} = -i\tr_\mathcal{S} \left\{N(\rho^\mathcal{S}_\tau)^{N-1}\tr_\mathcal{E} \left[ H_{tot},\rho^\mathcal{SE}_\tau\right]\right\}= -i\,N \tr_\mathcal{SE} \left\{ (\rho^\mathcal{S}_\tau)^{N-1}\otimes \I^\mathcal{E} \left[ H_{int},\rho^\mathcal{SE}_\tau\right]\right\}$, 
where the cyclic property of the trace was used. Note that the dependence on $H^\mathcal{S}$ and $H^\mathcal{E}$ vanishes \footnote{$H^\mathcal{S}$ vanishes since $\tr_\mathcal{SE}\left\{ \rho^\mathcal{SE} \left[\rho^\mathcal{S}\otimes \I^\mathcal{E},H^\mathcal{S} \otimes \I^\mathcal{E}\right] \right\} = \tr_\mathcal{S} \left\{ [\rho^\mathcal{S},H^\mathcal{S}]\tr_\mathcal{E} \rho^\mathcal{SE}  \right\}$. Using the cyclic property again, $\tr_\mathcal{S} \left\{ [\rho^\mathcal{S},H^\mathcal{S}]\tr_\mathcal{E} \rho^\mathcal{SE}  \right\}= \tr_\mathcal{S}\left\{ H^\mathcal{S}[\rho^\mathcal{S},\rho^\mathcal{S}]\right\} = 0$. The term $H^\mathcal{E}$ vanishes in a similar manner.}. Using the cyclic property of the trace once more, we find,$\left[ \frac{d}{dt}f_N\right]_{\tau} = \sum_j N\,p_j^{N-1}\;\left[\frac{d}{dt}p_j\right]_{\tau}=i\,N \tr\Big\{H_{int} \left[(\rho^\mathcal{S}_\tau)^{N-1}\otimes \I^\mathcal{E},\rho^\mathcal{SE}_\tau\right] \Big\}$.
With this result at hand, which is valid for all powers $N$, we do a series expansion of the von Neumann entropy to obtain the exact expression for the entropy rate:
\begin{equation}\label{eq:entropyrate}
\left[ \frac{d}{dt}\mathbb{S}\right]_{t=\tau} =  -i\,\tr_\mathcal{SE}\Big\{H_{int} \left[\ln(\rho^\mathcal{S}_\tau)\otimes \I^\mathcal{E},\rho^\mathcal{SE}_\tau\right] \Big\},  
\end{equation}
which expresses the dependence of the entropy rate in terms of the system-environment interaction $H_{int}$ and the commutator $\left[\ln(\rho^\mathcal{S})\otimes \I^\mathcal{E},\;\rho^\mathcal{SE}\right]$. This expression is true for any kind of system-environment couplings, including strong coupling and is highly non-Markovian. Note that $\left[\ln(\rho^\mathcal{S})\otimes \I^\mathcal{E},\;\rho^\mathcal{SE}\right]=0\Leftrightarrow \left[\rho^\mathcal{S}\otimes \I^\mathcal{E},\;\rho^\mathcal{SE}\right]=0$. We will use this commutator often in this Letter, and refer to it as:
\begin{equation}\mathfrak{C}\left(\rho^\mathcal{SE} \right):=\left[\rho^\mathcal{S}\otimes \I^\mathcal{E},\rho^\mathcal{SE}\right],
\end{equation}which contains some important properties of the structure of the total system-environment state with respect to its system part.
Considering the properties of the total $\mathcal{SE}$ state separate from the details of the Hamiltonian allow us to obtain the first important result of the paper:  
\begin{equation}\label{stable}\left[ \frac{d}{dt}\mathbb{S}_t\right]_{t=\tau} = 0\;\;\forall\; H_{tot}\quad \Leftrightarrow \quad \mathfrak{C}\left(\rho^\mathcal{SE} \right)=0.
\end{equation}
Eq.~(\ref{stable}) means that the necessary and sufficient condition for the entropy rate of the system to be zero under any coupling to the environment $H_{int}$ is that the bipartite state $\rho^\mathcal{SE}_\tau$ has the property $\mathfrak{C}\left(\rho^\mathcal{SE}\right)=0$. 

We refer to the $\mathcal{SE}$ states that have the property $\mathfrak{C}\left(\rho^\mathcal{SE}\right)=0$ as \emph{lazy states}. Lazy states do not have to be eigenstates of $H_{tot}$ nor of $H_{int}$. Effectively, they lead to closed system dynamics for very short times, independent of the details of the Hamiltonian coupling \footnote{This comes from $\sum_j p_j^N \dot{p}_j=0\; \forall\, N$. For the generic case where $\rho^\mathcal{S}$ is non-degenerate, this is equivalent to having first derivative of the spectrum vanish.}. The class of lazy states is also different from the concept of subdecoherent states and decoherence-free subspaces \cite{Zanardi97,Zanardi98,PhysRevLett.81.2594,Zanardi04,Roa07,PhysRevLett.89.080401,Zanardi99, PhysRevLett.82.2417}, as lazy states are independent from the particular symmetries of the dynamics.  Lazy states are a natural consequence of dynamical stability of the entropy measure under arbitrary open system dynamics, connecting the dynamical properties of reduced systems $\mathcal{S}$ to the structure of the total $\mathcal{SE}$ state \cite{ref:HKO,ref:Kimura}. When the lazy states condition is satisfied, the entropy of the system can in principle be preserved by fast measurements or dynamical decoupling techniques \cite{PhysRevLett.89.080401,Zanardi99, PhysRevLett.82.2417}. The connection between the condition $\frac{d}{dt}\mathbb{S} \vert_{t=\tau}=0$ and decoherence suppression is analogous to the connection between the Zeno time, where $\langle \psi_0 \vert\frac{d}{dt}\vert \psi_t \rangle  \vert_{0}= 0$, and the quantum Zeno effect \cite{misra:756,PhysRevLett.89.080401}.

When Eq.~(\ref{stable}) holds, the Markovian approximation is inadequate to model decoherence as it would imply that there is no decoherence. The standard assumption of the initial condition corresponding to a system uncorrelated from its environment, $\rho^\mathcal{SE}\approx\rho^\mathcal{S}\otimes\rho^\mathcal{E}$, corresponds to a lazy state, and its entropy does not change for short times. A non-zero entropy rate only occurs for non-lazy states. In a different context, Ferraro \emph{et al.}~\cite{ferraro09a} showed that lazy states are sparse in the space of density matrices, in both the sense of volume and topology, having measure zero in the whole Hilbert space and nowhere dense. This highlights the limitations of the physical requirements necessary to derive the Markovian quantum master equation \cite{Gorini76a}. 
So far, we have discussed how lazy states are connected to the specific dynamical properties of Eq.~(\ref{stable}). However, lazy states also have other important properties in terms of bipartite quantum correlations. In fact, lazy states can be thought of a generalization of classically-correlated states as defined by quantum discord \cite{ref:OZ, ref:HV,ref:OZ2,ref:HV2, Datta10}. Quantum discord is a useful measure that assigns a degree of quantumness to $\mathcal{SE}$ correlations. When the discord is zero, the state is said to have only classical correlations. Quantum discord quantifies the difference between the quantum mutual information of $\mathcal{S}$ and $\mathcal{E}$, and the mutual information after an optimal set of measurements $\{\vert j \rangle\langle j \vert^\mathcal{S}\otimes\I^\mathcal{E} \}$. A state $\rho^\mathcal{SE}$ is classically correlated (has zero discord) if and only if it has the form $\rho^\mathcal{SE} = \sum_j p_j \ketbra{j}{j}^\mathcal{S} \otimes \rho^\mathcal{E}_j$, where $\left\{ \ket{j} \right\}$ form a rank-1 orthonormal basis  of $\mathcal{S}$, $\{p_j\}$ are the corresponding probabilities, and $\rho^\mathcal{E}_j$ are density matrices. This classicality is equivalently expressed by an invariance under the set of measurements $\{\vert j \rangle\langle j \vert^\mathcal{S}\otimes\I^\mathcal{E} \}$ such that $\rho^\mathcal{SE} = \sum_j \vert j \rangle\langle j \vert^\mathcal{S}\otimes\I^\mathcal{E}  \rho^\mathcal{SE} \vert j \rangle\langle j \vert^\mathcal{S}\otimes\I^\mathcal{E}$ \footnote{See \cite{Rodriguez07a,PhysRevA.81.012313}, Lemma 8.12 in \cite{ref:Hayashi} and Theorem 2 in \cite{ref:Datta}.}. 

According to quantum discord, classical correlations are characterized by bipartite states that are invariant under a set of measurements given by rank-1 projectors on $\mathcal{S}$. We now show that by generalizing this concept to projectors of higher rank, we obtain the set of lazy states. That is, let $\rho^\mathcal{SE}$ be an arbitrary bipartite state on the $\mathcal{SE}$ space.
Then,
\begin{equation}\label{eq:PM}
\mathfrak{C}\left(\rho^\mathcal{SE}\right) =0\;\;\Leftrightarrow\;\; \rho^\mathcal{SE} = \sum_{j} \Pi_j^\mathcal{S} \otimes \I^\mathcal{E}  \rho^\mathcal{SE} \Pi_j^\mathcal{S} \otimes \I^\mathcal{E}, 
\end{equation}where $\rho^\mathcal{S} = \sum_j p_j \Pi_j^\mathcal{S}$ and $\{\Pi_j^\mathcal{S}\}$ are orthonormal projectors that span the space of $\mathcal{S}$, but need not be of rank-1 or of the same dimensionality such that $\{p_j\}$ are non-degenerate. 

To prove Eq.~(\ref{eq:PM}), assume that $\mathfrak{C}\left(\rho^\mathcal{SE}\right)= 0$. Since $\Pi_j^\mathcal{S} \otimes \I^\mathcal{E}$ is a  projector to the eigenspace of $\rho^\mathcal{S} \otimes \I^\mathcal{E}$, we have $\rho^\mathcal{SE} \Pi_j^\mathcal{S} \otimes \I^\mathcal{E} = \Pi_j^\mathcal{S} \otimes \I^\mathcal{E} \rho^\mathcal{SE}=\Pi_j^\mathcal{S} \otimes \I^\mathcal{E}  \rho^\mathcal{SE} \Pi_j^\mathcal{S} \otimes \I^\mathcal{E}$. By the completeness $\sum_{j} \Pi_j^\mathcal{S} \otimes \I^\mathcal{E} = \I^\mathcal{S} \otimes \I^\mathcal{E}$, we have that $\sum_{j} \Pi_j^\mathcal{S} \otimes \I^\mathcal{E} \rho^\mathcal{SE} = \rho^\mathcal{SE}$. The converse can be seen from direct calculation.

The right side of Eq.~\eqref{eq:PM} is the post-measurement state under local projectors $\{ \Pi_j\}$. These states include uncorrelated states, maximally entangled states and states with zero quantum discord, and any other states that satisfy $\mathfrak{C}=0$. These properties can be exploited to design experiments that detect global bipartite quantum correlations by monitoring only the dynamics of a subsystem. Connecting Eq.~(\ref{stable}) to Eq.~(\ref{eq:PM}) allows us to conclude that under the presence of any interaction with its environment, the entropy rate of a system is zero if the system has classical correlations, as defined by quantum discord. If the entropy of the local subsystem has a non-zero time derivative, then the total state is not a lazy state, and it has a non-zero quantum discord. This can serve as a protocol that detects quantum discord between $\mathcal{S}$ and $\mathcal{E}$ by monitoring the purity of $\mathcal{S}$, without knowledge of any environmental properties or of the total Hamiltonian. Previous results for uncorrelated states \cite{ref:GOH,ref:GOM} are special cases of this. 

Lazy states also provide a dynamical explanation for the robustness of measurement apparatus against decoherence. For finite-dimensional Hilbert space, the orthonormal states that define the measurement apparatus $\{\ketbra{\mu_i}{\mu_i}\}$ are called pointer states and uniquely specify the measured quantity. Eq.~(\ref{stable}) and Eq.~\eqref{eq:PM} provide a dynamical argument in favor of the stability of these pointer states \cite{RevModPhys.75.715}. Let $\mathcal{Q}$ be a quantum state to be measured and $\mathcal{M}$ be the macroscopic measurement apparatus. This is equivalent to relabeling $\mathcal{S}\rightarrow \mathcal{M}$ and $\mathcal{E}\rightarrow \mathcal{Q}$.  The act of a measurements correlates the measurement apparatus with the quantum state into $\rho^\mathcal{MQ} = \sum_i p_i\ketbra{\mu_i}{\mu_i}^\mathcal{M}\otimes \rho^\mathcal{Q}_i $.  This is a classically correlated state from the apparatus' point of view, and thus is a lazy states.

Furthermore, by maintaining the conceptual separation of the Hamiltonian dynamics from the structure of the total state we establish a new connection between non-lazy states and the entropy rate. Consider a bounded operator $A$ and a trace-class operator $\sigma$ on a Hilbert space. It follows that $\lvert \tr\left[A\sigma\right] \rvert \le \lVert A \sigma \rVert_1 \le \lVert A \rVert \lVert\sigma\rVert_1$ \cite{Schatten}. Applying this to Eq.~(\ref{eq:entropyrate}), we arrive to another important result: 
\begin{equation}\label{srate}
\biggl\lvert { \;\frac{d}{dt}\mathbb{S}^\mathcal{S}_t \;\biggr\rvert_{t=\tau}} \le \bigl\lVert H_{int}\bigr\rVert \; \bigl\lVert \left[\ln(\rho^\mathcal{S})\otimes \I^\mathcal{E},\;\rho^\mathcal{SE}\right]\bigr \rVert _1. 
\end{equation}
In Eq.~(\ref{srate}), $\lVert H_{int} \rVert$ defines the time-scale of the decoherence process, and $\lVert \left[\ln(\rho^\mathcal{S})\otimes \I^\mathcal{E},\;\rho^\mathcal{SE}\right]\rVert_1$ provides a universal bound on the rate of decoherence for a system-environment interaction $H_{int}$ of arbitrary strength. Thus, there is a maximum magnitude of entropy rate that come only from the structure of the total state.
This result can also be used to estimate the rate of decoherence, as measured by the entropy rate time $\tau$, from partial knowledge of the total $\mathcal{SE}$ state in relationship to $\mathcal{S}$ and partial information about the strength of the $\mathcal{SE}$ interaction.

To obtain some conceptual understanding of what the norm of the commutator in Eq.~(\ref{srate}) means, we will now focus on a simpler case. A similar exact result to Eq.~({\ref{eq:entropyrate}}) can be found for the purity (linear entropy) rate of the system: $\left[ \frac{d}{dt}\mathbb{P}\right]_{t=\tau} =  i\,\tr_\mathcal{SE}\left\{H_{int} \left[\rho^\mathcal{S}_\tau\otimes \I^\mathcal{E},\rho^\mathcal{SE}_\tau\right] \right\}$. 
Likewise, the magnitude of the entropy rate of the system is bound by: $\lvert { \;\frac{d}{dt}\mathbb{P}_t \;\rvert_{t=\tau}} \le  2 \lVert H_{int}\rVert \; \lVert \mathfrak{C}\left(\rho^\mathcal{SE}\right) \rVert _1$. The quantity $\lVert \mathfrak{C}\left(\rho^\mathcal{SE}\right) \rVert_1$ measures how ``far'' total system-environment density matrix is from commuting with the reduced system density matrix. To get some intuition for the meaning of $\lVert \mathfrak{C} \rVert_1$, we will now consider only $\mathcal{SE}$ states that are pure, $\ketbra{\chi}{\chi}^\mathcal{SE}$. We can show how this is not a restrictive class by invoking the Church of the Larger Hilbert Space \cite{Nielsen00a}. By defining an ancillary space $\mathcal{A}$, any state $\rho^\mathcal{SE}$ can be purified into $\ketbra{\chi}{\chi}^\mathcal{SEA}$ \cite{Stinespring}. Since we are interested in properties of the evolution of the system $\mathcal{S}$, we can refer to the rest of the Hilbert space $\mathcal{EA}$ simply as a new environment, and for simplicity relabel it as $\mathcal{EA}\rightarrow\mathcal{E}$. Simillarly, the total Hamiltonian can be thought of trivially acting on the ancilla, $H_{tot}\rightarrow H_{tot}\otimes\I^\mathcal{A}$.

It is easy to show that for pure states $\varrho^{\SA\EA} = \ketbra{\chi}{\chi}$ all the standard notions of uncorrelated states, classically correlated states, and separable (not entangled) states coincide. Using this fact, we can show that the purity rate is bounded by the quantum mutual information, the entropy of entanglement, \cite{PhysRevA.53.2046}, and the quantum discord. From \cite{ref:GOH}, $\lvert  \frac{d}{dt}\mathbb{P} \rvert \le 4 ||H_{int}|| \sqrt{2 I(\chi)}$, where $I(\chi):= S(\varrho^{\SA}) + S(\varrho^{\EA}) - S(\varrho^{\SA\EA})$ is the quantum mutual information. Since $\varrho^\mathcal{SE}$ is pure, it follows that $I(\chi) = 2 S(\varrho^\mathcal{S})  =2 E(\chi)= 2 \delta_{ \mathcal{S}\to \mathcal{E}}(\chi)$ where  $E(\chi):= S(\varrho^{\SA})$ is the entropy of entanglement and $\delta_{\mathcal{S}\to \mathcal{E}}(\chi)$ is the quantum discord. Thus $\lVert\, \mathfrak{C}\left( \ketbra{\chi}{\chi}\right) \,\rVert_1$ is bounded by the amount of $\mathcal{SE}$ correlations. Previous work \cite{Duer01} shows how this is related to the entanglement power of a Hamiltonian.

In addition, pure lazy states have very simple properties that allow us to connect them to other types of correlations. Let $\varrho^\mathcal{SE} = \ketbra{\chi}{\chi}$ be a pure state, with Schmidt decomposition $\chi = \sum_{i=1}^s \sqrt{p_i}\psi_i\otimes \phi_i \ (p_i > 0)$, where the elements of $\{\sqrt{p_i}\}$ are Schmidt coefficients and $s \le \min[d_s,d_e]$ is the Schmidt rank. 
Using $\varrho^\mathcal{S} = \sum_{i=1}^s p_i \ketbra{\psi_i}{\psi_i}$ to explicitly calculate $\mathfrak{C}\left(\varrho^\mathcal{SE}\right)  = 0$, we have $p_i^{\frac{3}{2}}p_k^{\frac{1}{2}} = p_k^{\frac{3}{2}}p_i^{\frac{1}{2}}$ for all $i,k = 1,\ldots,s$. This is satisfied if and only if $p_i = \frac{1}{s}$ for all $i = 1,\ldots, s$. We conclude that, $\varrho^\mathcal{SE} = \ketbra{\chi}{\chi}$ is a lazy state if and only if $\ p_i = \frac{1}{s}$. These states include the important class of maximally-entangled states. We can use this result to connect $\lVert \mathfrak{C} \rVert_1$ to robustness of entanglement \cite{ref:Rob} and the negativity \cite{ref:Neg}. By direct computation one has $\mathfrak{C}_\chi:=\mathfrak{C}\left(\varrho^\mathcal{SE}\right) = \sum_{i\neq k} \sqrt{p_i p_k}(p_i-p_k) \ketbra{\psi_i}{\psi_k}\otimes \ketbra{\phi_i}{\phi_k}$. 
 Taking the trace norm and using the triangle inequality, 
we obtain $\mathfrak{C}_\chi \le  \sum_{i\neq k} \sqrt{p_i p_k}|p_i-p_k|$. The inequality is generally strict but if the Schmidt number is $2$, the equality holds. Since $0 \le p_i \le 1 $, one has $|p_i-p_k| \le 1$, which gives the bound: $\mathfrak{C}_\chi \le \sum_{i \neq k} \sqrt{p_i p_k} = (\sum_i \sqrt{p_i})^2 - 1$. 
The right hand side is the robustness of entanglement $R(\chi)$ for pure states \cite{ref:Rob}, which coincides with $2 N(\chi)$, where $N(\chi)$ is the negativity of $\chi$ \cite{ref:Neg}. Here we have shown how for pure states $\lVert \mathfrak{C} \rVert_1$ is bound by the total amount of correlations.

In conclusion, we focus on studying universal properties of the system dynamics that are independent of the details of the system-environment coupling, but that depend on the structure of the total system-environment state.  We defined the class of lazy states, which have the property $\left[\rho^\mathcal{S}\otimes \I^\mathcal{E},\rho^\mathcal{SE}\right]=0$, and prove that this is a sufficient and necessary condition for the system entropy rate to be zero. This effectively makes the open system dynamics to act as if they were closed dynamics for short times. Lazy states are a generalization of the classically-correlated states, as defined by quantum discord. This result was used to explain the dynamical stability of a measurement apparatus and the pervasive nature of decoherence. Also, we proposed an experimental protocol for detecting global quantum correlations from local observables. Finally, we showed how the time-derivative of the purity is bounded by the amount of system-environment correlations, establishing that bipartite correlations not only restrict the entropy of a subsystem, but also its rate of change. 

Future work will include a generalization of lazy states for continuous variables. Also, further exploration is needed to understand $\lVert \mathfrak{C} \rVert_1$. Can it be thought as a distance measure to lazy states? Work along these lines might reveal other features of decoherence dynamics that are universal for any Hamiltonian.


We thank K. Modi, K. Imafuku, J. Whitfield, M. Mosonyi, G. Sudarshan, H. Ohno, S. Boixo, P. Love and W. Zurek and  for fruitful discussions. This material is based upon work supported as part of the Center for Excitonics, an Energy Frontier Research Center funded by the U.S. Department of Energy, Office of Science, Office of Basic Energy Sciences under Award Number DE-SC0001088 and by the Office of Naval Research under Award Number N000140911049.

\bibliography{discordderivative.bib}
\end{document}